\documentclass[reprint,superscriptaddress,amsmath,amssymb,aps,prl,floatfix]{revtex4-2}

\usepackage{graphicx}
\usepackage{dcolumn}
\usepackage{bm}
\usepackage[output-decimal-marker={.},exponent-product=\cdot]{siunitx}
\usepackage{makecell}
\usepackage{color}
\usepackage[english]{babel}
\usepackage[T1]{fontenc}
\usepackage{multirow}

\usepackage{xcolor}
\usepackage[colorlinks = true, citecolor=blue,urlcolor=blue,linkcolor=blue]{hyperref}
\usepackage{url}
\usepackage{mathtools, cuted}
\usepackage{booktabs}
\usepackage[mathlines]{lineno}
\begin{document}

\title{Ultra-low $Q_\beta$ value for the allowed decay of $^{110}$Ag$^m$ confirmed via mass measurements}

%% Authors
\author{J.~Ruotsalainen}
\email{jouni.k.a.ruotsalainen@jyu.fi}
\affiliation{University of Jyvaskyla, Department of Physics, Accelerator laboratory, P.O. Box 35(YFL) FI-40014 University of Jyvaskyla, Finland}
\author{M.~Stryjczyk}
\email{marek.m.stryjczyk@jyu.fi}
\affiliation{University of Jyvaskyla, Department of Physics, Accelerator laboratory, P.O. Box 35(YFL) FI-40014 University of Jyvaskyla, Finland}
\author{M. Ramalho}
\email{madeoliv@jyu.fi}
\affiliation{University of Jyvaskyla, Department of Physics, Accelerator laboratory, P.O. Box 35(YFL) FI-40014 University of Jyvaskyla, Finland}
\author{T.~Eronen}
\affiliation{University of Jyvaskyla, Department of Physics, Accelerator laboratory, P.O. Box 35(YFL) FI-40014 University of Jyvaskyla, Finland}
\author{Z.~Ge}
\affiliation{University of Jyvaskyla, Department of Physics, Accelerator laboratory, P.O. Box 35(YFL) FI-40014 University of Jyvaskyla, Finland}
\author{A.~Kankainen}
\affiliation{University of Jyvaskyla, Department of Physics, Accelerator laboratory, P.O. Box 35(YFL) FI-40014 University of Jyvaskyla, Finland}
\author{M.~Mougeot}
\affiliation{University of Jyvaskyla, Department of Physics, Accelerator laboratory, P.O. Box 35(YFL) FI-40014 University of Jyvaskyla, Finland}
\author{J. Suhonen}
\email{jouni.t.suhonen@jyu.fi}
\affiliation{University of Jyvaskyla, Department of Physics, Accelerator laboratory, P.O. Box 35(YFL) FI-40014 University of Jyvaskyla, Finland}
\affiliation{International Centre for Advanced Training and Research in Physics (CIFRA), P.O. Box MG12, 077125 Bucharest-M\u{a}gurele, Romania}

\begin{abstract}
%The mass of $^{109}$Ag isotope was measured with respect to $^{110}$Cd with the phase-imaging ion-cyclotron-resonance technique using the JYFLTRAP double Penning trap at the IGISOL facility. By combining this new result with the known spectroscopic information, the $Q_\beta$ value of $405(135)$~eV between the $6^+$ isomer of $^{110}$Ag and the $5^+_2$ state in $^{110}$Cd was determined. This is the lowest $Q_\beta$ for any allowed transition observed to date. The nuclear shell-model results were performed with the {\sc kshell} code employing the \textit{jj45pnb} interaction, and combined with the state-of-the-art atomic calculations. The theoretical partial half-life, $2.23^{+5.24}_{-1.28} \times 10^7$ years, and the resulting branching ratio, $3.07^{+4.16}_{-2.15} \times 10^{-8}$, combined with a viable production method, thermal-neutron capture on stable $^{109}$Ag, make $^{110}$Ag$^{m}$ as a promising candidate for future antineutrino mass measurement experiments.
%%%% modified 2024-11-22 
The mass of the electron-antineutrino can be determined in dedicated measurements of the $\beta$ spectral shape near the $\beta$ endpoint of a $\beta^-$ transition, with a low $Q$ value enhancing the sensitivity of the measurement. One such low-$Q$-value candidate is the transition between the $6^+$ isomer of $^{110}$Ag and the $5^+_2$ state in $^{110}$Cd ($Q^{\ast}_{\beta,m}=-0.12(131)$ keV). To reduce the uncertainty of the $Q$ value, we have used the phase-imaging ion-cyclotron-resonance technique with the JYFLTRAP double Penning trap and performed a high-precision atomic-mass measurement of $^{109}$Ag with $^{110}$Cd as a reference. Combined with the known spectroscopic data, we obtain a re-evaluated value $Q^{\ast}_{\beta,m}=405(135)$ eV, for the $^{110}\text{Ag}(6^+_\text{m}) \rightarrow {^{110}\text{Cd}}(5^+_2)$ transition. This represents the lowest $Q_{\beta}$ value for any allowed transition observed to date. In order to estimate the partial half-life ($t_{1/2}$) and branching ratio (Br) of the transition, nuclear shell-model calculations were performed using the $jj45pnb$ Hamiltonian in combination with state-of-the-art atomic calculations. The computed values $t_{1/2} = 2.23^{+5.24}_{-1.28} \times 10^7$ years and $\textrm{Br} = 3.07^{+4.16}_{-2.15} \times 10^{-8}$, along with the thermal-neutron capture on stable $^{109}$Ag as a viable production method, make $^{110}\textrm{Ag}^m$ a promising candidate for future antineutrino-mass measurements.
\end{abstract}

\maketitle
The determination of the absolute mass of neutrinos remains one of the big unanswered questions in physics. Currently, the most stringent limits come from the kinematic experiments: KATRIN \cite{Aker2021,Aker2022} and MARE \cite{Nucciotti2008,Ferri2015} for the electron antineutrino, and ECHo \cite{Gastaldo2017,Velte2019} and HOLMES \cite{Faverzani2016,DeGerone2022} for the electron neutrino. They are based on the model-independent analysis of the end-point spectrum of $\beta^-$ decays of $^{3}$H (KATRIN) and $^{187}$Re (MARE) or $^{163}$Ho electron capture (ECHo and HOLMES). Another tritium-based experiment, Project 8, is currently being planned \cite{Pettus2020}. These three isotopes have been chosen due to the very small ground-state-to-ground-state $Q_\beta$ values (\num{18592.071(22)}~eV for $^{3}$H \cite{MedinaRestrepo2023}, \num{2863.2(6)}~eV for $^{163}$Ho \cite{Schweiger2024} and \num{2470.9(13)}~eV for $^{187}$Re \cite{Filianin2021}). Nevertheless, despite such narrow energy windows, only a very small fraction of decays ends up near the endpoint, limiting the sensitivity of these experiments. 
In neutrino mass experiments involving ${\beta}$ decay and electron capture (EC), a low $Q$ value enhances sensitivity to (anti)neutrino mass. Thus, finding such decays with the smallest possible $Q$ value is desirable for experiments aimed at determining the (anti)neutrino mass.

An alternative approach has been recently proposed where the decay proceeds to the excited nuclear state in the daughter isotope with an ultra-low $Q^*$ value, usually ${<1}$~keV  \cite{Kopp2010}. The first promising $\beta^-$-decay candidate discovered is $^{115}$In where the second-forbidden unique decay to the first excited states in $^{115}$Sn ($9/2^+_{\text{gs}}\rightarrow 3/2^+_1$) was experimentally observed \cite{Wieslander2009,Andreotti2011} and its $Q_\beta^*$ value is around 200~eV \cite{Mount2009,Wieslander2009}. Two other $\beta^-$-decay isotopes of interest are $^{135}$Cs where $Q_\beta^* = 0.44(31)$~keV for the first-forbidden unique decay to the excited $11/2^-$ state  \cite{deRoubin2020} as well as $^{131}$I with the $Q_\beta^*$ value of $1.03(23)$~keV for an allowed decay to the excited $9/2^+$ state. \cite{Eronen2022}.
Additionally, a promising EC candidate $^{159}$Dy was recently discovered \cite{Ge2021a} with the $Q_{EC}^*$ of 1.18(19) keV for the $^{159}$Dy(3/2$^-$) $\rightarrow$ $^{159}$Tb$^*$(5/2$^-$) transition. However, with the exception of $^{115}$In, these ultra-low-$Q^*$ decays are still awaiting experimental confirmation.

While $^{115}$In seems to be an ideal candidate for a new generation of neutrino-mass experiments, a small branching ratio ($\approx 10^{-6}$) resulting in a very long partial half-life ($\approx 10^{20}$~y \cite{Wieslander2009}) is limiting its use. As a result, the search for other cases continues \cite{Sandler2019,Karthein2019,Ge2021,Ge2021a,Ramalho2022,Ge2022,Ge2022a,Gamage2022,Gamage2022a,Ge2023,Ge2024,Ge2024a}, with an emphasis on allowed transitions as they are the fastest and their $\beta$ shape can be described analytically \cite{Keblbeck2023}. However, these studies are focused only on ground-state-to-excited-state  decays \cite{Keblbeck2023}, disregarding possible isomeric-state-to-excited-state (iso-ex) transitions.
 
\begin{figure}[!htb]
   \centering
   \includegraphics[width=\columnwidth]{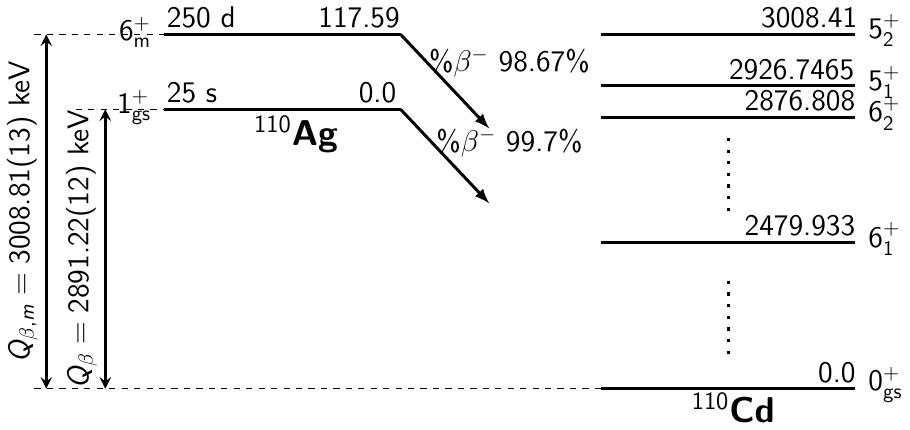}
   \caption{\label{fig:decayscheme}The $\beta^-$ decay scheme of $^{110}$Ag with the states in $^{110}$Cd relevant for this work, based on Refs. \cite{Gurdal2012,110CdXUNDL}. The $Q_\beta$ values are from this work.}
\end{figure}

One isomer that could be of interest for future antineutrino mass experiments is $^{110}$Ag$^{m}$ ($E_m = 117.59(5)$~keV, $J^\pi = 6^+$, ${T_{1/2} \approx 250}$~d \cite{NUBASE2020,Gurdal2012}).
It decays predominantly (98.67(8)\%) via $\beta^-$ to the excited states in stable $^{110}$Cd (see Fig.~\ref{fig:decayscheme}), with a minor decay branch (1.33(8)\%) via internal transition \cite{NUBASE2020,Gurdal2012}. Additionally, it can be readily produced in nuclear reactors due to the large cross section ($\approx 4.1$~b) for thermal-neutron capture on stable $^{109}$Ag \cite{Nakamura2003,Elmaghraby2019}. The literature $Q_\beta(^{110}\text{Ag}^{m}) \equiv Q_{\beta,m}$ value is 3008.3(13)~keV \cite{AME2020,NUBASE2020} and the recent studies constrained the excitation energy of the non-isomeric $^{110}\text{Cd} (5^+_2)$ state to $E_x = 3008.41(4)$~keV \cite{110CdXUNDL}, resulting in the present energy window for the allowed iso-ex decay of $Q_{\beta,m}^* = -0.12(131)$~keV. 
Interestingly, hints of the $5^+_2$ state being populated in the $\beta^-$-decay of $^{110}$Ag$^{m}$ have already been observed. The 846 keV $\gamma$ ray, which is the strongest transition de-exciting the $5^+_2$ state \cite{110CdXUNDL} was detected in two studies  \cite{Mallet1981,Kiang1993}. In addition, a second transition de-exciting this level, the 1466 keV $\gamma$ ray, was observed in one of these studies  \cite{Mallet1981}. However, both of them remain unplaced in the  $\beta^-$-decay of $^{110}$Ag \cite{Mallet1981,Kiang1993,Gurdal2012}.
All these facts make $^{110}$Ag$^{m}$ a perfect candidate for a potential ultra-low-$Q_\beta^*$ transition. 

%-----------
%--------------

The masses of $^{109}$Ag and $^{110}$Ag isotopes are connected via a precise single-neutron separation energy value from the $(n,\gamma)$ study ($S_n = 6809.19(11)$~keV \cite{Bogdanovic1981}) while the $^{110}$Ag$^{m}$ and $^{110}\text{Cd}(5^+_2)$ excitation energies are known from $\gamma$-ray spectroscopy studies \cite{Gurdal2012}. As a result, $Q_{\beta,m}^*$ can be expressed as a function of masses of two stable isotopes, $^{109}$Ag and $^{110}$Cd, and precise spectroscopic observables. Considering that the quadratically-added uncertainties of $S_n$, $E_m$ and $E_x$ are equal to 127~eV, the $\approx1.3$~keV uncertainty of $Q_{\beta,m}^*$ \cite{AME2020,NUBASE2020} can be reduced if the single-proton separation energy ($S_p$) of $^{110}$Cd is known with a precision better than 100 eV. In order to accomplish it, the mass of $^{109}$Ag was determined with $^{110}$Cd as a reference by utilizing the JYFLTRAP double Penning trap mass spectrometer \cite{Eronen2012} at the Ion Guide Isotope Separator On-Line (IGISOL) facility at the University of Jyv\"askyl\"a \cite{Moore2013}. This measurement results in the most precise determination of the $Q^*_{\beta,m}$, which is presented in this letter. To evaluate the potential of this transition for neutrino mass detection, we incorporated the precise $Q_{\beta,m}^*$ value into nuclear shell-model calculations using the KSHELL code with the jj45pnb interaction, combined with advanced atomic calculations. 
Additionally, to check for possible mass-dependent systematic effects caused by the $m(^{110}\text{Cd})-m(^{109}\text{Ag})\approx 1$~u mass difference, two mass measurements of  zirconium oxide molecules, $^{90}$Zr$^{16}$O and $^{92}$Zr$^{16}$O with $^{91}$Zr$^{16}$O as a reference, were also conducted, see Supplemental Material \cite{supplemental}. 
These molecules are in the same mass region as $^{109}$Ag with nuclear masses that are well characterized  \cite{Lone1979,Baglin2013,AME2020,NUBASE2020}.

To produce the ions used in the measurements, a sample of natural silver was placed in an electric discharge ion source located in the target area of the IGISOL \cite{Moore2013}, while samples of natural cadmium and zirconium were placed in the offline discharge ion source station \cite{Vilen2020}. The extracted ions were accelerated by a 30~kV electrostatic potential. The ions were transported through a 55$^{\circ}$ dipole magnet capable of separating the isobars containing the desired ions, $^{109}$Ag$^+$, $^{110}$Cd$^+$ and $^{90-92}$Zr$^+$, from the ion beam and allowing them to enter the radio-frequency quadrupole cooler-buncher \cite{Nieminen2001}. In addition to cooling and bunching the beam for more efficient injection into the JYFLTRAP, the ZrO$^+$ molecules were formed there. Subsequently, the bunches were injected into the helium-filled preparation trap of JYFLTRAP \cite{Eronen2012}, where the mass-selective buffer-gas cooling technique \cite{Savard1991} was applied to separate the ions of interest from possible isobaric contaminants. The ions of interest were then sent to the measurement trap of JYFLTRAP. 

In the measurement trap, the phase-imaging ion-cyclotron-resonance (PI-ICR) technique \cite{Eliseev2013,Eliseev2014,Nesterenko2018,Nesterenko2021} was used to measure the free space cyclotron frequency $\nu_c=qB/(2\pi m)$ of an ion species, where $q$ and $m$ are the charge and mass of the ion and $B$ the magnetic field strength. By measuring the cyclotron frequencies of an ion of interest ($\nu_{c,ioi}$) and a reference ion ($\nu_{c,ref}$), we obtain the ratio of the mass of the ion of interest to the mass of the reference from the ratio $r$ of their frequencies, as
\begin{equation}
    r=\frac{\nu_{c,ref}}{\nu_{c,ioi}}=\frac{m_{ioi}}{m_{ref}}.
\end{equation}

\begin{figure*}
   \centering
   \includegraphics[width=\textwidth]{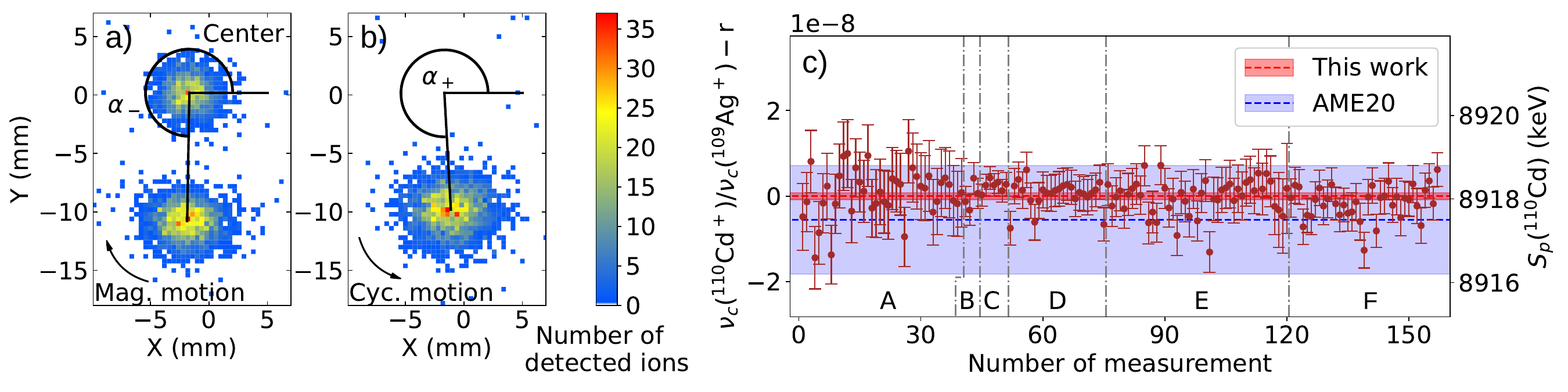}
   \caption{\label{fig:piicr}A PI-ICR measurement of a) magnetron and center spots and b) a cyclotron spot of $^{109}$Ag$^+$ ions collected on the 2D-MCP detector after 2.5 hours with $t_\text{acc} = 1$~s. c) Measured frequency ratios $\nu(^{110}$Cd$^+$)/$\nu(^{109}$Ag$^+$) with the gray dotted lines indicating breaks in the measurement and the accumulation time change (A: $t_\text{acc} = 1$~s, B: $t_\text{acc} = 1$~s, C: $t_\text{acc} = 1.5$~s, D: $t_\text{acc} = 1$~s, E: $t_\text{acc} = 813$~ms and F: $t_\text{acc} = 1$~s). For the dataset E, the magnetron and reduced cyclotron spots were collected at $\alpha_-\approx\alpha_+\approx 0^\circ$ while for other datasets $\alpha_-\approx\alpha_+\approx 270 ^\circ$. The average frequency ratio $r = $~\num{0.99091693576(44)} and the corresponding $S_p$ value for $^{110}$Cd are indicated with the red band while the AME20 value is plotted in blue.}
\end{figure*}

With the PI-ICR technique, $\nu_c$ is determined by projecting the ions in the trap onto a position-sensitive microchannel plate detector (2D-MCP). The cyclotron frequency is obtained using the angle between the centers of the spots formed on the detector relative to the center spot after magnetron ($\alpha_-$) and reduced cyclotron ($\alpha_+$) eigenmotion accumulation
\begin{equation}
    \nu_c=\frac{(\alpha_+-\alpha_-)+2\pi n_c}{2\pi t_{acc}}\mathrm{.}
    \label{eq:frequency}
\end{equation}
where $t_\text{acc}$ is the accumation time and $n_c$ the number of revolutions during $t_\text{acc}$ at frequency $\nu_c$. Two timing patterns were used to excite the ions' radial eigenmotions using radiofrequency electric fields in order to produce the spots. At the start of both patterns, a dipolar pulse with frequency $\nu_-$ was used to reduce the initial magnetron motion of the ions to bring them to the center of the trap. Then for the magnetron spot, the pattern continued with a dipolar radiofrequency pulse with frequency $\nu_+$ exciting the reduced cyclotron motion. This was followed by a quadrupolar pulse with frequency $\nu_c$ converting the motion to magnetron motion. Finally, after waiting for an accumulation time $t_\text{acc}$, the ions were projected onto the 2D-MCP registering their final positions. For the reduced cyclotron spot, the order of the converting pulse and the accumulation time was reversed. The position of the center of the trap on the 2D-MCP was determined by extracting the centered ions without further excitations. Typical position data obtained in this work are shown in Fig.~\ref{fig:piicr}a) and b). 

For the mass determination of $^{109}$Ag, a total of 157 frequency ratios between $^{109}$Ag$^+$ and the reference $^{110}$Cd$^+$ were measured over 40 hours using three different accumulation times, 813 ms, 1 s and 1.5 s, see Fig.~\ref{fig:piicr}c). Two different spot positions on the detector were used to account for distortion in the projected image, but no systematic shift in the measured ratio was observed. A final ratio $r$ was obtained as a weighted average of all the measurements. Similarly for $^{92}$Zr$^{16}$O$^+$ mass determination, 98 ratios over 25 hours were measured using two different spot positions and two accumulation times of 813 ms and 1 s, while for $^{90}$Zr$^{16}$O$^+$, 20 ratios with $t_\text{acc} = 1$~s were collected over five hours, both against the $^{91}$Zr$^{16}$O$^+$ reference, see Supplemental Material~\cite{supplemental}.

To reduce the effect of residual magnetron and cyclotron motion on the ions' position, the start of the conversion pulse was scanned over the cyclotron period ($\sim$ 1 $\mu$s) and the extraction from the measurement trap over the period of the magnetron motion ($\sim$ 600 $\mu$s). The effect of the temporal magnetic field fluctuation was minimized by alternating the measurements of the ion of interest and the reference ion every seven minutes and linearly interpolating the frequency $\nu_c$ of the reference at the time of the ion of interest measurement. The temporal fluctuation of the JYFLTRAP magnetic field has been measured to be $\delta B/(B\delta t) = 2.01\times10^{-12}$ min$^{-1}$ \cite{Nesterenko2021}. By performing count-rate class analysis \cite{Kellerbauer2003,Roux2013,Nesterenko2021} using bunches with up to five detected ions, the effect of ion-ion interactions was determined to be negligible. 

The single-nucleon separation energy $S_x$ can be determined from the frequency ratio $r$ measured between two neighboring species:
\begin{equation}
S_x = \big[-|r-1|\times(m_{\text{ref}} - m_e) + m_x\big]c^2 + \frac{1-r}{|1-r|}(rB_\text{ref} - B_\text{ioi})\mathrm{,}
\end{equation} 
where $m_{\text{ref}}$ and $m_e$ are the atomic mass of the reference species and the mass of an electron, respectively, while $B_\text{ref}$ and $B_\text{ioi}$ are the first ionization energies of the reference ion and the ion of interest, respectively, taken from Ref. \cite{NIST}. For a single-neutron separation energy, $m_x$ is the atomic mass of a neutron and for a single-proton separation energy $m_x$ represents the mass of a hydrogen atom.

\begin{table}
\centering
\caption{\label{tab:sx} The frequency ratios $r$ determined in this work with respect to the reference ions (Ref.) and the corresponding single-nucleon separation energies $S_{x}$ (proton for $^{110}$Cd, neutron for $^{91,92}$Zr) compared to the literature values $S_{x,\text{lit.}}$ from Refs. \cite{AME2020,Baglin2013,RzacaUrban2018}.}
\begin{ruledtabular}
\begin{tabular}{lllll}
Nuclide     & Ref.          & $r=\nu_{c,\text{ref}}/\nu_{c}$                       & $S_{x}$ (keV)    & $S_{x,\text{lit.}}$ (keV)  \\\hline
$^{109}$Ag  & $^{110}$Cd    & $0.99091693576(44)$   & 8918.07(5)            & 8917.5(13) \\ 
$^{92}$Zr$^{16}$O & $^{91}$Zr$^{16}$O  & $1.00934887819(52)$ & 8634.74(5) & 8634.75(4) \\ 
\multirow{2}{*}{$^{90}$Zr$^{16}$O} & \multirow{2}{*}{$^{91}$Zr$^{16}$O}  & \multirow{2}{*}{$0.9906366606(12)$} & \multirow{2}{*}{7194.75(12)} & 7194.36(15)\footnotemark[1] \\ 
& & &  & 7194.9(4)\footnotemark[2] \\ 
\end{tabular}
\end{ruledtabular}
\footnotetext[1]{From AME20 \cite{AME2020}.}
\footnotetext[2]{From the ENSDF evaluation \cite{Baglin2013} based on the $(n,\gamma)$ study from Ref. \cite{Lone1979}.}
\end{table}

The measured frequency ratios and corresponding single-particle separation energies are presented in Table~\ref{tab:sx}. The $^{92}$Zr single-neutron separation energy perfectly agrees with the Atomic Mass Evaluation 2020 (AME20) which is based on the $(n,\gamma)$ study from Ref. \cite{RzacaUrban2018}. At the same time, the $S_n(^{91}\text{Zr})$ value differs by 0.38(19)~keV from AME20. However, it agrees with a value from the $(n,\gamma)$ measurement \cite{Lone1979} evaluated at ENSDF \cite{Baglin2013}. Considering that between the ion of interest and the reference ion there was only one mass unit of difference, the same as between $^{109}$Ag and $^{110}$Cd, and that the ZrO molecules are in the same mass region, no additional mass-dependent uncertainties are added to the $^{109}$Ag frequency ratio.  

The updated proton separation energy, $S_p(^{110}\text{Cd}) = 8918.07(5)$~keV, and the mass excess of $^{109}$Ag (ME = \num{-88718.8(4)}~keV)  are consistent with the values reported in AME20 ($S_{p,\text{lit.}}(^{110}\text{Cd}) = 8917.5(13)$~keV, ME$_\text{lit.}$ = \num{-88719.4(13)}~keV \cite{AME2020}). These updated values are 26 and three times more precise, respectively, with the uncertainty in the latter being dominated by the uncertainty in the mass of $^{110}$Cd.
These results allows for reevaluation of the $Q_{\beta,m}$ value:
\begin{equation}
\begin{split}
Q_{\beta,m} &= \big[m(^{110}\mathrm{Ag}^{m}) - m(^{110}\mathrm{Cd})\big]c^2 \\&= \big[(r-1)\times(m(^{110}\mathrm{Cd}) - m_e) + m_n\big]c^2 \\&+ E_m - S_n + (rB_{\mathrm{Cd}}- B_{\mathrm{Ag}})\mathrm{,}
\end{split}
\end{equation} 
where $r = \num{0.99091693576(44)}$ is the frequency ratio between $^{110}$Cd$^+$ and $^{109}$Ag$^+$ from this work, $m(^{110}\mathrm{Cd}) = \num{109903007.5(4)}$~$\mu$u is the atomic mass of $^{110}$Cd \cite{AME2020}, $m_e$ and $m_n$ are the masses of an electron and a neutron, respectively, $E_m = 117.59(5)$~keV \cite{NUBASE2020} and $S_n = 6809.19(11)$~keV \cite{Bogdanovic1981} are the $^{110}$Ag$^{m}$ excitation energy and the $^{110}$Ag single-neutron separation energy, respectively, while $B_{\mathrm{Cd}} = 8.993820(16)$~eV and $B_{\mathrm{Ag}} = 7.576234(25)$~eV are the first ionization energies of Cd and Ag, respectively \cite{NIST}.

\begin{table}
\centering
\caption{\label{tab:qbeta} A comparison of $Q_\beta$ values between different states in $^{110}$Ag and $^{110}$Cd from AME20 \cite{AME2020} and this work. The $^{110}$Ag$(6^+_\text{m})$ and $^{110}$Cd$(5^+_2)$ excitation energies are taken from Refs. \cite{NUBASE2020,110CdXUNDL}.}
\begin{ruledtabular}
\begin{tabular}{lll}
Transition                      & AME20                     & This work  \\\hline
$1^+_{\text{gs}}\rightarrow 0^+_{\text{gs}}$  & \num{2890.7(13)} keV      & \num{2891.22(12)} keV \\ 
$6^+_\text{m}\rightarrow 0^+_{\text{gs}}$     & \num{3008.3(13)} keV      & \num{3008.81(13)} keV \\ 
$6^+_\text{m}\rightarrow 5^+_2$        & \num{-0.12(131)} keV      & \num{405(135)} eV \\
\end{tabular}
\end{ruledtabular}
\end{table}

All the $\beta$-decay energies extracted in this work agree with AME20 \cite{AME2020} but they are an order of magnitude more precise, see Table~\ref{tab:qbeta}. 
The updated $Q_{\beta,m}^*$ value confirms that the transition is energetically allowed at a confidence level ($Q/\delta Q$) of $3\sigma$, with ${Q_{\beta,m}^* = 405(135)}$~eV. 
It is the lowest observed $Q_{\beta}^*$ value for an allowed transition and the second-lowest overall after the second-forbidden unique decay of $^{115}$In to the first excited state in $^{115}$Sn \cite{Keblbeck2023}. 

To estimate the partial half-lives for the decays of $^{110}$Ag to the excited states in $^{110}$Cd, we utilized the {\sc kshell} code \cite{Shi2019} within the Nuclear Shell Model (NSM) framework, employing the \textit{jj45pnb} interaction \cite{Lisetskiy2004}, a two-nucleon potential based on a perturbative G-matrix approach. The model space includes the proton orbitals \(0f_{5/2}\), \(1p_{3/2}\), \(1p_{1/2}\), and \(1g_{9/2}\) with single-particle energies of $-14.938$, $-13.437$, $-12.044$, and $-8.905$~MeV, respectively. For neutrons, the model space comprises the \(0g_{7/2}\), \(1d_{5/2}\), \(1d_{3/2}\), \(2s_{1/2}\), and \(0h_{11/2}\) orbitals, with corresponding single-particle energies of $6.230$, $2.442$, $2.945$, $2.674$, and $4.380$~MeV, as specified in the \textit{jj45pnb} interaction. The computations were performed without truncation of the model space. 
We note that for the nuclear mass region around $A=110$, only variants of the $jj45pnb$ Hamiltonian are available in the libraries of the shell-model codes KSHELL and NuShellX@MSU \cite{Brown2014}. All these variants produce similar results so that we consider here the results obtained by using the $jj45pnb$ Hamiltonian as representative of all these Hamiltonians.

\begin{table*}%[h!t!b]
\caption{\label{tab:allowed_decays}Comparison of experimental (column 2) and computed (columns 3-6) partial half-lives $t_{1/2}$ for allowed decays of \({}^{110}\)Ag to \({}^{110}\)Cd calculated with different values of the effective weak-axial coupling (\(g_{{\rm A}}^{{\rm eff}}\)). The NSM-computed Gamow-Teller matrix element ($M_{\rm GT}$) and the experimental decay energies ($Q_\beta^*$) utilized in the computations are also presented. Experimental data was gathered from Refs. \cite{Gurdal2012,110CdXUNDL} and this work.}
\begin{ruledtabular}
\begin{tabular}{lcccccccc}
   \multirow{2}{*}{Transition}  & \multicolumn{6}{c}{$t_{1/2}$} & \multirow{2}{*}{$M_{\rm GT}$} & \multirow{2}{*}{$Q_\beta^*$ (keV)} \\\cmidrule(lr){2-7}
    & Exp. & 0.5 & 0.6 & 0.7 & 0.8 & Units & & \\\hline
    \(6_{\text{m}}^+ \phantom{a} \rightarrow \phantom{a} 6^+_1\) & $808.51^{+8.06}_{-7.90}$ & 561.61 & 390.007 & 286.536 & 219.379 & d & 0.0117 & 528.88(13) \\
    \(6_{\text{m}}^+ \phantom{a} \rightarrow \phantom{a} 5^+_1\)  & $369.02^{+9.81}_{-4.24}$ & 117.627 & 81.6857 & 60.0139 & 45.9482 & d & -0.4742 & 82.07(13) \\
    \(6_{\text{m}}^+ \phantom{a} \rightarrow \phantom{a} 5^+_2\)  & $-$\footnotemark[1] & 22.309 & 15.492 & 11.382 & 8.714 & My & 0.1542 & 0.405(135) \\
    \(1^+_{\text{gs}} \phantom{a} \rightarrow \phantom{a} 0^+_{\text{gs}}\)  & $25.88^{+0.20}_{-0.20}$ & 21.85 & 15.18 & 11.15 & 8.54 & s & 0.7271 & 2891.22(12) \\
\end{tabular}
\end{ruledtabular}
\footnotetext[1]{This transition has not been observed experimentally yet.}
\end{table*}

To validate the applied interaction, we computed the excitation energies, magnetic dipole and electric quadrupole moments for both ${}^{110}$Ag and ${}^{110}$Cd, and compared them with evaluated data, see Supplemental Material~\cite{supplemental}. 
In computations of the electromagnetic moments, we have used standard effective charges $e^p_{\rm eff}=1.5e$ for protons and $e_{\rm eff}^n = 0.5e$ for neutrons, and the bare g-factors $g_l(p) = 1$, $g_l(n) = 0$, $g_s(p) = 5.585$, and $g_s(n) = -3.826$. This choice leads to a reasonable agreement between the computed and measured electromagnetic moments, indicating that the wave functions of the involved nuclear states are fairly reliable. The computed excitation energies of the $6^+_m$ and $5^+_2$ states, involved in the ultra-low-$Q_{\beta}$ transition, correspond rather well to their experimental counterparts, thus speaking for their reliability in the half-life and branching estimations.

The major known allowed $\beta$-decay transitions from $^{110}$Ag to $^{110}$Cd as well as the transition of interest were computed and compared with experimental results, see Table~\ref{tab:allowed_decays}. These calculations included screening, radiative, and atomic exchange corrections, originally derived for allowed $\beta$ decays and shown to have a stronger contribution at low electron energies \cite{Nit2023}. Additionally, the effective weak axial coupling (\(g_{{\rm A}}^{{\rm eff}}\)) was varied between 0.5 and 0.8 as suggested in Ref.~\cite{Ejiri2019} for this mass region. 

\begin{figure}[!htb]
   \centering
   \includegraphics[width=\columnwidth]{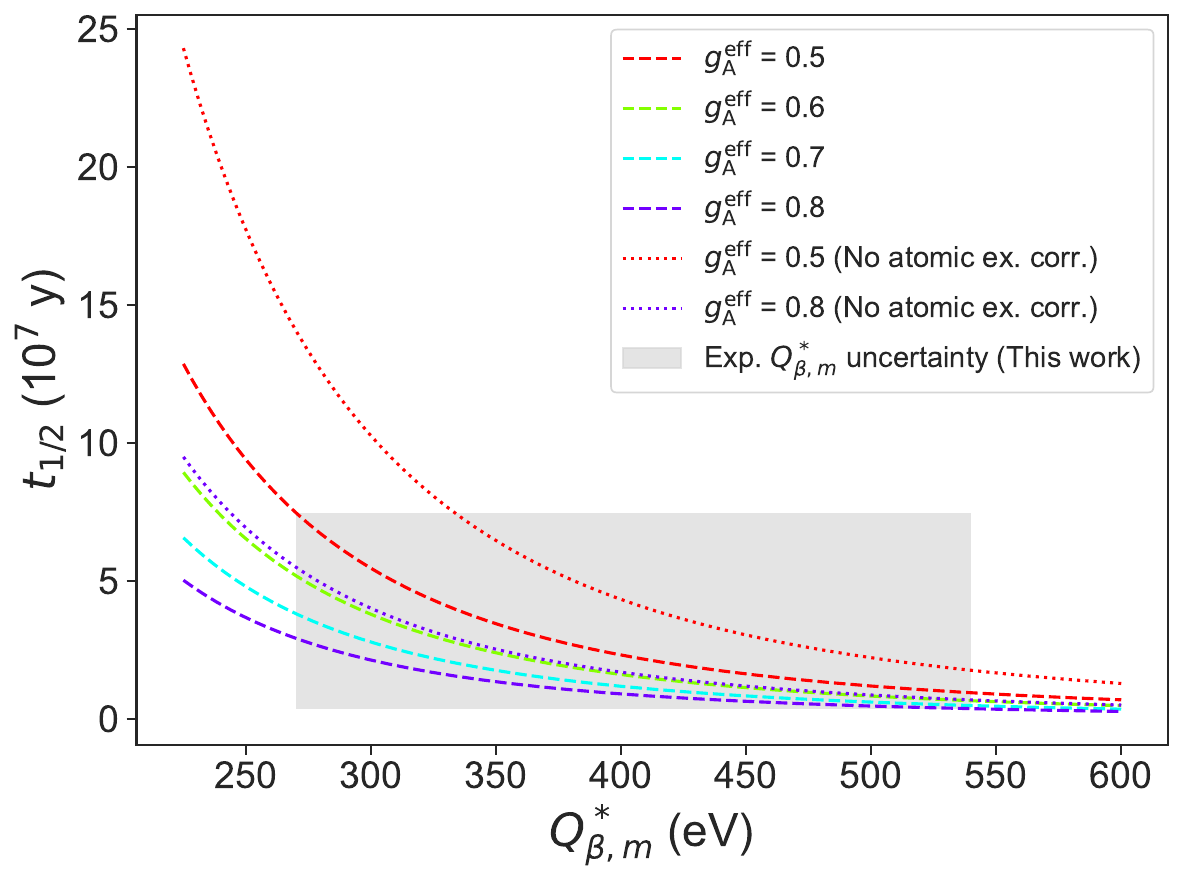}
   \caption{Theoretical estimates of the partial half-life $t_{1/2}$ of the allowed \(^{110}{\rm Ag}(6^+_\text{m})\rightarrow\,^{110}{\rm Cd}(5_{2}^+)\) transition with a $Q_{\beta,m}^*$ value of 405(135) eV for different effective weak-axial coupling values (\(g_{{\rm A}}^{{\rm eff}}\)). The gray shaded area represents the $1\sigma$ uncertainty of the $Q_{\beta,m}^*$ value from this work. The dotted lines represent the computations without the atomic exchange correction.}
   \label{fig:Decays}
\end{figure}

Since the phase-space factor is extremely sensitive to the decay energy, for the $6^+_\text{m}\rightarrow 5^+_2$ transition the expected half-lives for the different \(g_{{\rm A}}^{{\rm eff}}\) values and for the $Q_{\beta,m}^*$ energies ranging from 225 to 600 eV are plotted, see Fig.~\ref{fig:Decays}. The experimental $1\sigma$ uncertainties found in this work are indicated by a gray shaded region. The partial half-life $t_{1/2}$ ranges from \(3.70 \times 10^{6}\)~y for the upper $Q_{\beta,m}^*$ limit (540 eV) with \(g_{{\rm A}}^{{\rm eff}}=0.80\) up to \(7.47 \times 10^{7}\)~y for the lower $Q_{\beta,m}^*$ limit (270 eV) with \(g_{{\rm A}}^{{\rm eff}}=0.5\). For any given \(g_{{\rm A}}^{{\rm eff}}\) curve, we mostly observe an order of magnitude difference in $t_{1/2}$ whilst considering the gray shaded area. The partial half-lives for two \(g_{{\rm A}}^{{\rm eff}}\) values, 0.5 and 0.8, were also computed without incorporating the atomic exchange corrections \cite{Nit2023} to assess its impact. As can be seen in Fig.~\ref{fig:Decays}, the $t_{1/2}$ values increase by a factor $\approx1.9$, and they range between \(6.88 \times 10^{6}\) and \(1.41 \times 10^{8}\) years. This result highlights the importance of this correction and emphasizes the importance of high-precision measurements of the ultra-low $Q_\beta^*$ for the half-life determination. The observed discrepancies between theoretical predictions and experimental half-lives can be linked to limitations of the Nuclear Shell Model. These include the exclusion of orbitals outside the defined model space, the neglect of nuclear deformation effects, and the lack of fine-tuning of interaction parameters for this isotope. While addressing these issues would require significant effort, the results still serve as a valuable benchmark for evaluating the current computational framework.

The theoretical half-lives for the strongest $\beta$-decay branches are best reproduced when $g_{{\rm A}}^{{\rm eff}} = 0.5$, as shown in Table~\ref{tab:allowed_decays}. Therefore, this value was employed for the calculations concerning the main transition of interest. This choice is motivated entirely by its better overall agreement with experimental results, although it reflects a broader overreach of the nuclear shell model, suggesting that the model might not fully capture the quenching effects observed in experimental data. To estimate the influence of the energy window, the experimental $Q_{\beta,m}^*$ value was varied by $1\sigma$. The resulting partial half-life is $2.23^{+5.24}_{-1.28} \times 10^7$ years, which corresponds to a branching ratio of $3.07^{+4.16}_{-2.15} \times 10^{-8}$ for the $^{110}\text{Ag}(6^+_\text{m}) \rightarrow {^{110}\text{Cd}}(5^+_2)$ decay.

In summary, we have measured with high precision the single-proton separation energy of $^{110}$Cd, $S_p = 8918.07(5)$~keV, resulting in the improved mass-excess value of $^{109}$Ag, $\text{ME} = -88718.8(4)$~keV. By combining our results with the well known single-neutron separation energy and the isomer excitation energy of $^{110}$Ag, we demonstrated that the $^{110}\text{Ag}$ $(6^+_\text{m})\rightarrow {^{110}\text{Cd}^*}$ $(5^+_2)$
%$^{110}\text{Ag}(6^+_\text{m})\rightarrow {^{110}\text{Cd}}(5^+_2)$ 
$\beta$-decay transition has the ultra-low $Q^*_\beta$ value of 405(135)~eV. This is the lowest $Q^*_\beta$ for any allowed transition measured to date and the first low $Q^*_\beta$ case measured for a $\beta$-decaying isomer. The nuclear shell model combined with the atomic calculations predicted the partial half-life for this transition to be $2.23^{+5.24}_{-1.28} \times 10^7$ years which translates to the branching ratio of  $3.07^{+4.16}_{-2.15} \times 10^{-8}$. It should be noted that if the 846 keV and 1466 keV $\gamma$ rays observed in the decay studies of $^{110}$Ag originate from the de-excitation of the $5^+_2$ state in $^{110}$Cd, the partial decay half-life to this state would be in the $10^3-10^4$~years range, which is orders of magnitude shorter than indicated by calculations. All in all, our work indicates that $^{110}$Ag$^{m}$ might be a good candidate for the future antineutrino mass measurement experiments, although a new decay study of $^{110}$Ag$^{m}$ is required to confirm the production of the $5^+_2$ state in $^{110}$Cd. Our work also demonstrates that the future search of low-$Q_\beta^*$ transitions should not be limited to the $\beta$-decaying ground states, with the isomers in $^{101,102}$Rh, $^{106}$Ag, $^{120}$Sb, $^{121}$Te, $^{184}$Re and $^{194}$Ir being the most promising cases to find such transitions.

\begin{acknowledgments}

This project has received funding from the European Union’s Horizon 2020 research and innovation programme under grant agreement No. 771036 (ERC CoG MAIDEN), from the European Union’s Horizon Europe Research and Innovation Programme under Grant Agreement No. 101057511 (EURO-LABS) and the Research Council of Finland projects No. 295207, 306980, 327629, 354589 and 354968. J.R. acknowledges the financial support from the Vilho, Yrj\"o and Kalle V\"ais\"al\"a Foundation.

\end{acknowledgments}

\bibliography{mybibfile}

\end{document}

%title is a placeholder, to be changed to something nicer
\title{Supplemental material for "Ultra-low $Q_\beta$ value for the allowed decay of $^{110}$Ag$^m$ confirmed via mass measurements"}

\maketitle

\begin{figure}[!htb]
   \centering
   \includegraphics[width=\columnwidth]{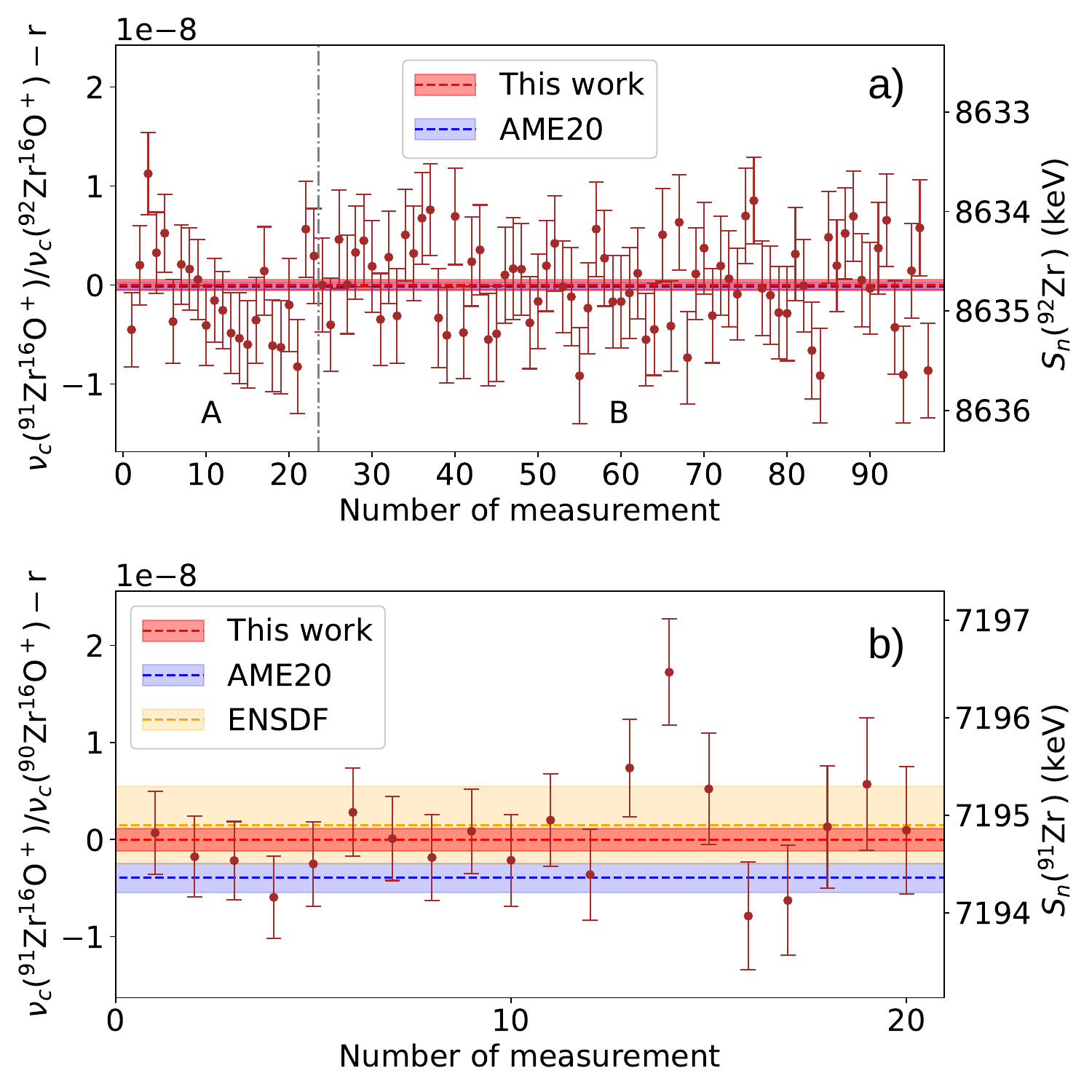}
   \caption{a) Measured frequency ratios $\nu(^{91}\text{Zr}^{16}\text{O}^+)/\nu(^{92}\text{Zr}^{16}\text{O}^+)$ with the gray dotted lines indicating breaks in the measurement and an accumulation time change (A: $t_\text{acc} = 1$~s, B: $t_\text{acc} = 813$~ms). The magnetron and cyclotron spots for dataset A were collected at an angle of $\alpha_+\approx\alpha_-\approx90^{\circ}$ and for dataset B $\alpha_+\approx\alpha_-\approx0^{\circ}$. The average frequency ratio $r = $~\num{1.00934887819(52)} and the corresponding $S_n$ value for $^{92}$Zr are indicated with the red band while the AME20 value \cite{AME2020} is plotted in blue. b) Measured frequency ratios $\nu(^{91}\text{Zr}^{16}\text{O}^+)/\nu(^{90}\text{Zr}^{16}O^+)$ with accumulation time $t_\text{acc} = 1$~s with the magnetron and cyclotron spots at an angle of $\alpha_+\approx\alpha_-\approx 270^{\circ}$. The average frequency ratio $r = $~\num{0.9906366606(12)} and the corresponding $S_n$ value for $^{91}$Zr are indicated with the red band, the AME20 value \cite{AME2020} is plotted in blue while the ENSDF value \cite{Baglin2013} based on Ref. \cite{Lone1979} in yellow.}
\end{figure}

\begin{table}[!htb]
\caption{\label{tab:EM_props}Comparison of the electric quadrupole ($Q_s$) and magnetic dipole ($\mu$) moments with those computed by the NSM using the \textit{jj45pnb} Hamiltonian. The experimental values are taken from Refs. \cite{Gurdal2012,Stone2016,Stone2019,Stone2020}.}
\begin{ruledtabular}
\begin{tabular}{lccccc}
    \multirow{2}{*}{Nuclide} & \multirow{2}{*}{\( J^{\pi} \)} & \multicolumn{2}{c}{Experimental} & \multicolumn{2}{c}{Computed} \\
    \cmidrule(lr){3-4} \cmidrule(lr){5-6}
    & & $Q_s$ ($eb$) & $\mu$ ($\mu_N$) & $Q_s$ ($eb$) & $\mu$ ($\mu_N$) \\\hline
    \({}^{110}\)Ag & \(1^{+}_{\text{gs}}\) & 0.24(12) & 2.7225(14) & 0.1602 & 2.6072 \\
    \({}^{110}\)Ag & \(6^{+}_\text{m}\) & 1.44(10) & 3.602(4) & 0.8785 & 3.787 \\
    \({}^{110}\)Ag & \(3^{+}_1\) & -- & 3.77(3) & 0.5815 & 3.8243 \\
    \({}^{110}\)Cd & \(2^{+}_1\) & $-0.40(4)$ & 0.81(6) & $-0.32$ & 1.0851 \\
\end{tabular}
\end{ruledtabular}
\end{table}
% \vspace{2cm}

\begin{table}[!htb]
\caption{\label{tab:Ex110Ag}Comparison of the experimental ($E_{x}^\text{exp}$) and theoretical ($E_{x}^\text{th}$) excitation energies of the positive-parity states in $^{110}$Ag. The experimental values are taken from Refs. \cite{Gurdal2012,Zhang2023}.}
\begin{ruledtabular}
\begin{tabular}{ccc}
    $J^{\pi}$ & $E_{x}^\text{exp}$ (keV) & $E_{x}^\text{th}$ (keV) \\\hline
    $1^{+}_{\text{gs}}$ & 0 & 0  \\
    $6^{+}_\text{m}$ & 117.59(5) & 207 \\
    $3^{+}_1$ & 118.719(10) & 20 \\
    $3^{+}_2$ & 191.622(12) & 319 \\
    $2^{+}_1$ & 198.689(10) & 259 \\
    $4^{+}_1$ & 266.7\footnotemark[1] & 72  \\
    $5^{+}_1$ & 379.1\footnotemark[1] & 178 \\
    $6^{+}_2$ & 446.89\footnotemark[1] & 276 \\
    $4^{+}_2$ & 471.1\footnotemark[1] & 275 \\
\end{tabular}
\end{ruledtabular}
\footnotetext[1]{Exact uncertainties are not given in the reference but are below 1 keV.}
\end{table}
% \vspace{2cm}

\begin{table}[h]
\caption{\label{tab:Ex110Cd}Comparison of the experimental ($E_{x}^\text{exp}$) and theoretical ($E_{x}^\text{th}$) excitation energies of the positive-parity states in $^{110}$Cd. The experimental values are taken from Refs. \cite{Gurdal2012,110CdXUNDL}.}
\begin{ruledtabular}
\begin{tabular}{ccc}
    $J^{\pi}$ & $E_{x}^\text{exp}$ (keV) & $E_{x}^\text{th}$ (keV) \\\hline
    $0^{+}_{\text{gs}}$ & 0 & 0  \\
    $2^{+}_1$ & 657.7623(11) & 611 \\
    $0^{+}_2$ & 1473.07(3) & 1493 \\
    $2^{+}_2$ & 1475.7900(14) & 1546 \\
    $4^{+}_1$ & 1542.4441(14) & 1583 \\
    $0^{+}_3$ & 1731.31(3) & 1758 \\
    $2^{+}_3$ & 1783.496(15) & 1926 \\
    $4^{+}_2$ & 1809.48(9) & 2205 \\
    $0^{+}_4$ & 2078.80(5) & 2369 \\
    $3^{+}_1$ & 2162.8015(15) & 1899 \\
    $4^{+}_3$ & 2220.0683(14) & 2331 \\
    $4^{+}_4$ & 2250.554(11) & 2526 \\
    $2^{+}_4$ & 2287.63(11) & 2027 \\
    $3^{+}_2$ & 2433.248(25) & 2056 \\
    $6^{+}_1$ & 2479.9339(25) & 2651 \\
    $5^{+}_1$ & 2926.7474(16) & 2556 \\
    $5^{+}_2$ & 3008.41(4) & 2782 \\
\end{tabular}
\end{ruledtabular}
\end{table}
% \vspace{4.5cm}

\bibliography{mybibfile}